\documentclass[fleqn,usenatbib]{mnras}
\usepackage{newtxtext,newtxmath}
\usepackage[T1]{fontenc}

\DeclareRobustCommand{\VAN}[3]{#2}
\let\VANthebibliography\thebibliography
\def\thebibliography{\DeclareRobustCommand{\VAN}[3]{##3}\VANthebibliography}
\usepackage{graphicx}	
\usepackage{amsmath}	

\def\bea{\begin{eqnarray}}
\def\eea{\end{eqnarray}}
\def\be{\begin{equation}}
\def\ee{\end{equation}}

\newcommand{\de}{\mathrm d}
\newcommand{\om}{{\Omega_{\rm m}}}
\newcommand{\ho}{{H_0}}

\title[RSB via angular correlations]{Constraints on the origin of the radio synchrotron background via angular correlations}

\author[E. Todarello et al.]{Elisa Todarello,$^{a,b}$\thanks{E-mail: elisamaria.todarello@unito.it}
    Marco Regis,$^{a,b}$\thanks{E-mail: marco.regis@unito.it}
    Federico Bianchini,$^{c}$
    Jack Singal,$^{d}$
    Enzo Branchini,$^{e}$
    \newauthor{Fraser J. Cowie,$^{f}$
   Sean Heston,$^{g}$
    Shunsaku Horiuchi,$^{g,h}$
    Danielle Lucero,$^{g}$
   Andre Offringa$^{i,j}$}
   \\  \\ 
$^{a}$ Dipartimento di Fisica, Università di Torino, Via P. Giuria 1, 10125 Torino, Italy\\
$^{b}$ Istituto Nazionale di Fisica Nucleare, Sezione di Torino, Via P. Giuria 1, 10125 Torino, Italy\\
$^{c}$ Kavli Institute for Particle Astrophysics and Cosmology, Stanford University, 382 Via Pueblo Mall, Stanford, CA 94305-4060, USA\\
$^{d}$ Physics Department, University of Richmond, 138 UR Drive, Richmond, VA 23173, USA\\
$^{e}$ Department of  Physics, University of Genova, Via Dodecaneso, 33, Genoa, Italy\\
$^{f}$ Astrophysics, Department of Physics, University of Oxford, Denys Wilkinson Building, Keble Road, Oxford OX1 3RH, UK\\
$^{g}$ Department of Physics, Virginia Tech University, Blacksburg, VA 24061-0435, USA\\
$^{h}$ Kavli IPMU (WPI), UTIAS, The University of Tokyo, Kashiwa, Chiba 277-8583, Japan\\
$^{i}$ Netherlands Institute for Radio Astronomy (ASTRON), 
Oude Hoogeveensedijk 4, 7991 PD Dwingeloo, Netherlands\\
$^{j}$ Kapteyn Astronomical Institute, 
P.O. Box 800, 9700 AV Groningen, Netherlands\\
}

\begin{document}
\label{firstpage}
\pagerange{\pageref{firstpage}--\pageref{lastpage}}
\maketitle

\begin{abstract}
    The origin of the radio synchrotron background (RSB) is currently unknown. Its understanding might have profound implications in fundamental physics or might reveal a new class of radio emitters. In this work, we consider the scenario in which the RSB is due to extragalactic radio sources and measure the angular cross-correlation of LOFAR images of the diffuse radio sky with matter tracers at different redshifts, provided by galaxy catalogs and CMB lensing.  We compare these measured cross-correlations to those expected for models of RSB sources. We find that low-redshift populations of discrete sources are excluded by the data, while higher redshift explanations are compatible with available observations.  We also conclude that at least 20\% of the RSB surface brightness level must originate from populations tracing the large-scale distribution of matter in the universe, indicating that at least this fraction of the RSB is of extragalactic origin.   Future measurements of the correlation between the RSB and tracers of high-redshift sources will be crucial to constraining the source population of the RSB.
\end{abstract}

\section{Introduction}
\label{sec:intro}

The background level of radio sky brightness, dominated by steep-spectrum synchrotron-like emission below 5 GHz and so here referred to as the radio synchrotron background (RSB), is due to some as of now unknown combination of the integrated emission from extragalactic radio sources and a possible large-scale Galactic halo,   $T_\mathrm{RSB}=T_\mathrm{G}+T_\mathrm{E}$.  Direct measurement of the level of the radio background, combining Absolute Radiometer for Cosmology, Astrophysics, and Diffuse Emission 2 (ARCADE~2) measurements from 3 to 90$\mathrm{~GHz}$ (\cite{Fixsen:2009xn}) with several radio maps at lower frequencies from which an absolute zero level has been inferred  (recently summarized in~\cite{Singal:2022jaf}), reveals a brightness spectrum (\cite{2018ApJ...858L...9D}):
\be
T_\mathrm{RSB}(\nu) = 30.4 \pm 2.6 \mathrm{K} \, \left(\frac{\nu}{310\mathrm{~MHz}}
\right)^{-2.66 \pm 0.04}\;.
\label{T_B}
\ee
At $\nu=140$ MHz, this radio background amounts to $T_\mathrm{RSB}\sim 252$ K. 
As summarized in, e.g. \cite{Singal:2022jaf}, it has been apparent for nearly 15 years that this level of surface brightness is several times higher than can be accounted for by known populations of extragalactic radio sources and accepted Galactic emission characteristics.

To estimate the Galactic component we consider \cite{Fornengo:2014mna}, and from their best-fits, we obtain $T_\mathrm{E}\sim 180$ K. 
The contribution from known sources down to a threshold of $\sim$5 mJy (which is the one of the images used in this work) is $\sim 30$ K, and can be obtained by integrating, e.g., the $dN/dS$ curve provided in \cite{Intema:2017}.

Therefore, if the RSB is of extragalactic origin, there remains a contribution from below-threshold sources of $T_\mathrm{bts}\sim 150$ K at 140 MHz, which is much higher than typical extrapolations for known radio populations.  We will refer to this unexplained contribution to the surface brightness as the ``RSB excess''.\footnote{Note that throughout this paper ``RSB excess'' refers to the excess surface brightness not accounted for by known source classes and not to any level of anisotropy power relative to some model.}  The observational statistical error on this estimate is small, of order of 10\%, whilst there are possible significant systematic errors related to the estimate of the Galactic contribution or to the zero-level calibration of the observations. Discussing those issues is not the goal of this work, and we will normalize the radio background contribution from extragalactic sources below 5 mJy to provide $T_\mathrm{bts}= 150$ K.

It is natural to expect that extragalactic sources providing the RSB would follow the matter distribution in the Universe. Their radio emission should therefore exhibit a certain level of correlation with cosmological matter tracers.  Analyses of the anisotropy angular power spectrum of the radio background include \cite{Offringa:2021rwp} and \cite{Cowie:2023rwp} at the same background radio frequencies considered in this analysis, and \cite{Holder14} at much higher background frequencies (above 4 GHz).  The latter work indicated that the observed anisotropy power of the radio background was lower than that which could result from sources following the large-scale distribution of matter in the universe, while the former works indicated, in contrast, an {\it excess} of anisotropy power above that which would be produced by known classes of extragalactic radio sources.

In this work, we constrain the properties of radio source populations explaining the RSB excess by computing their cross-correlation with the CMB lensing and with different galaxy catalogs. We measure the angular cross-correlation of LOFAR images at 140 MHz~\citep{Offringa:2021rwp,Cowie:2023rwp} with the {\it Planck} CMB lensing~\citep{carron22}, and the 2MPZ~\citep{Bilicki:2013sza}, SDSS main (Data Release 12)~\citep{alam15} and SDSS quasar (Data Release 16)~\citep{lyke20} catalogs.  By comparing those measurements to a simple but flexible phenomenological model of extragalactic radio emitters, we derive bounds on the redshift behavior required for a population that successfully provides the RSB excess.

The cross-correlation between the radio sky and low-$z$ galaxy catalogs has been discussed before in \citep{Brown:2010,Vernstrom:2017jvh}, even though with different goals than here and with shallower images from the Bonn survey~(\cite{Brown:2010}) and Murchison Widefield Array~\citep{Vernstrom:2017jvh}.

The paper is organized as follows. In Section~\ref{sec:obs}, we describe the observational data that we are going to use to perform the cross-correlation measurements. Section~\ref{sec:mod} introduces the phenomenological models adopted throughout the paper to describe extra-galactic radio sources.
The analysis, results, and discussion related to the angular autocorrelation, cross-correlation with CMB lensing, and cross-correlation with galaxy catalogs are provided in Sections~\ref{sec:auto}, \ref{sec:lens} and \ref{sec:gal}, respectively.
In Section~\ref{sec:con}, we summarize and discuss our findings.
In Appendix~\ref{sec:tomo}, we use the web tool Tomographer to further validate our results regarding the radio-catalog cross-correlation.

\section{Observational Data}
\label{sec:obs}

\begin{figure*}
\centering
\includegraphics[width=0.9\textwidth]{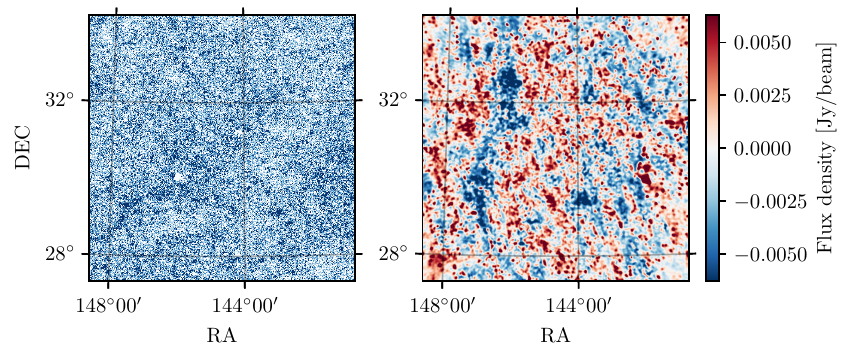}
\caption{Scatter plot of the SDSS main sample (left) in the field of the LOFAR coldest patch map (right).} 
\label{fig:coldest}
\end{figure*}

We consider radio maps from LOFAR observations obtained in high-band antenna dual mode.  Two of the maps were the result of a targeted observation at 142~MHz used for the analysis of Ref.~\cite{Offringa:2021rwp}.
The maps are centered at $9h\ 38m\ 41s,\ 30^\circ \ 49^\prime\ 12^{\prime\prime}$ (referred to as coldest patch) and $10h\ 25m\ 00s,\ 30^\circ \ 00^\prime\ 00^{\prime\prime}$ (referred to as reference field). Each image has $2500^2$ pixels and covers an area of approximately 48~deg$^2$. In these maps, bright sources with a flux larger than 7 times the rms fluctuation (i.e., with a flux $\gtrsim 5$~mJy) were selected in a uniformly weighted image that was tapered to make the synthesised beam a Gaussian
with a full width at half maximum (FWHM) of 30'' resolution. These sources have been removed from the visibilities. The residuals were then imaged with natural weighting, resulting in a synthesised beam that has a FWHM of $\sim 3.5$~arcmin.
Further details on observation and data reduction can be found in~\cite{Offringa:2021rwp}.  We also consider another map, coming from multi-beaming observations from legacy LOFAR data and being a mosaic composed by 7 adjacent fields, as described in \cite{Cowie:2023rwp}. The mosaic area is a $16\times 16$ deg$^2$, centered on the North Celestial Pole (NCP).
The synthesised beam has a size of approximately $3.3$~arcmin.
Sources have been subtracted off, as for the other images, again up to a threshold of around 5 mJy.
The images are naturally weighted to enhance sensitivity to the angular power spectrum.

\begin{figure*}
\centering
\includegraphics[width=0.9\textwidth]{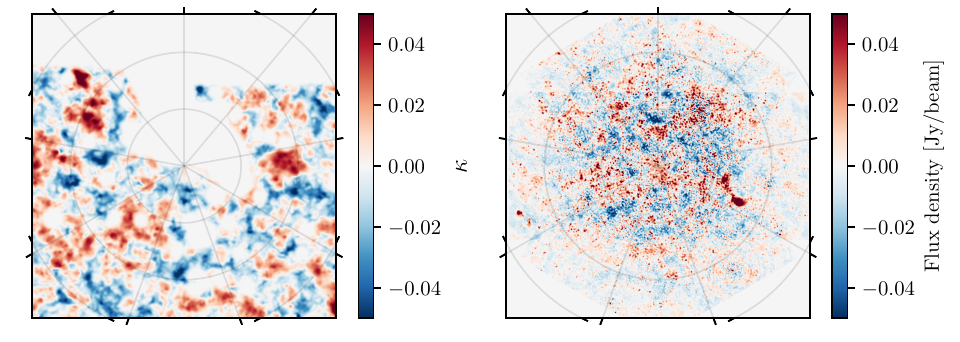}
\caption{A $16^\circ\times 16^\circ$ patch of sky centered on the North Celestial Pole as seen by the Widefield LOFAR High-Band Antenna at a radio frequency of 140 MHz (right panel), juxtaposed with a Wiener-filtered CMB lensing convergence map derived from the \textit{Planck} PR4 dataset (left panel, dimensionless). Gridlines of constant declination are shown every 3 degrees.}
\label{fig:cmb1}
\end{figure*}

\begin{figure*}
\centering
\includegraphics[width=0.48\textwidth]{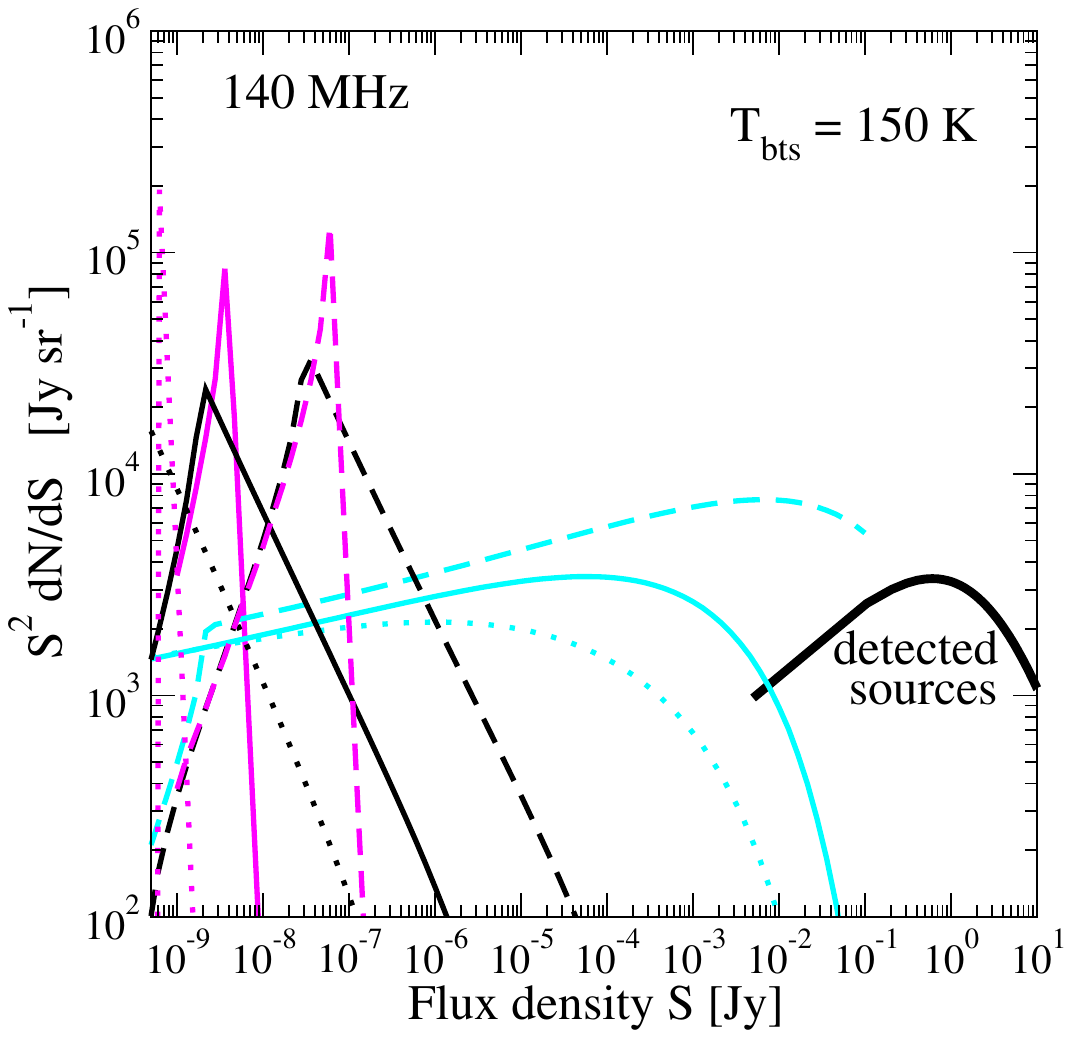}
\includegraphics[width=0.48\textwidth]{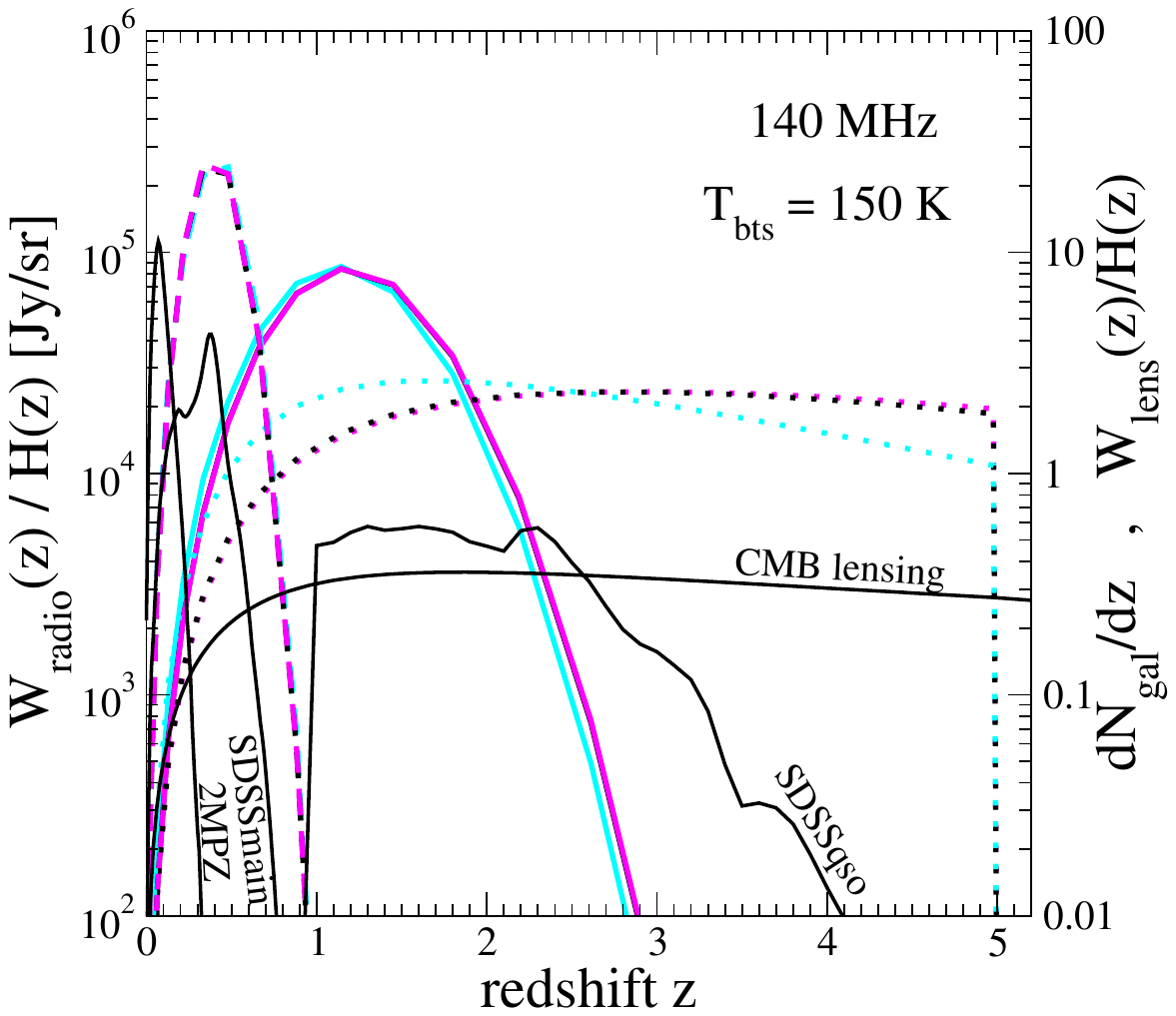}
\caption{Left: Illustrative examples of radio source count models ($dN/dS$) considered in this work, computed through Eq.~\eqref{eq:dNdS} and corresponding to models arising from Eq.~\eqref{eq:S} with weak, medium and strong scalings with the mass by taking $\alpha$ equal to 0.1 (magenta), 0.5 (black) and 1 (cyan), and for populations with distribution in redshift peaked at low or intermediate $z$, or flat in $z$, by taking $(z_0,\sigma)$ equal to (0.3, 0.15) (dashed), (1, 0.5) (solid), and (0,+$\infty$) (dotted). 
The bold black line at high fluxes shows the number counts measured from detected sources at 150 MHz~\protect\cite{Intema:2017}.
Right: Window functions of radio emitters (same color/style coding of the left panel, $y$-axis labels on the left); 
of galaxy/quasar catalogs and 
of CMB lensing (black solid curves, 
$y$-axis labels on the right).
}
\label{fig:dNdS}
\end{figure*}

As tracers of the large-scale structure, we consider three galaxy catalogs. The Two Micron All Sky Photometric Redshift catalog (2MPZ)~\citep{Bilicki:2013sza} is peaked at very low redshift, $z_{p}\simeq 0.07$, and contains $\sim 10^3$ galaxies in the region of a LOFAR field. 
The Sloan Digital Sky Survey (SDSS) main sample (Data Release 12)~\citep{alam15} has a wider distribution peaked at intermediate redshift, $z_{p}\simeq 0.35$, and offers higher statistics since the number of galaxies in the region of a LOFAR field amounts to  $\sim 10^5$. The content of the SDSS main sample in the coldest patch field is shown in Figure~\ref{fig:coldest}.
Finally, we consider the SDSS quasar catalog (Data Release 16)~\citep{lyke20}, so to cover higher redshifts. To this aim, namely, in order to test the correlation of the radio images with a population of sources at high redshift, we further remove sources with $z <1$ from the quasar catalog. In the region of a single LOFAR field, it then contains about $8\times 10^3$ objects.
The redshift distribution of the sources for the different catalogs is shown on the right panel of Figure~\ref{fig:dNdS}. To perform the angular cross-correlation measurement between a catalog and a radio image, we will actually consider sources from the catalog that are located within an area four times larger than the image, sharing the same center.

We also consider the publicly available cosmic microwave background (CMB) lensing convergence map $\kappa^{\rm CMB}$, which is reconstructed from \textit{Planck} PR4 data\footnote{\url{https://github.com/carronj/planck\_PR4\_lensing}} \citep{carron22}.
This lensing convergence map is extracted using an improved lensing quadratic estimator applied to the reprocessed  \textit{Planck} PR4 \texttt{NPIPE} foreground-cleaned CMB temperature and polarization maps.
These \texttt{NPIPE} maps incorporate about 8\% more data than the previous 2018 \textit{Planck} PR3 release.
For the CMB lensing reconstruction, CMB angular scales over the $100 \le \ell \le 2048$ range are optimally filtered and fed to the quadratic estimator.
We show in the left panel of Fig.~\ref{fig:cmb1} the Wiener-filtered \textit{Planck} CMB lensing convergence map over a patch of the sky centered on the North Celestial Pole.

\section{Phenomenological Models of Radio Source Populations}
\label{sec:mod}
We model the contribution of radio sources by assuming they populate halos in the Universe. The halo distribution is described adopting the halo mass function $\de n/(\de M\de V)$ from \cite{Sheth:1999mn}.
Then for each halo, we define an average radio flux density which depends on the halo mass and on the redshift. For simplicity, we take the two dependencies to be separable, i.e., the average radio flux density at 140 MHz of a halo of mass $M$ at redshift $z$ is $\mathcal{S}(M,z)=\mathcal{S}_0\,f(M)\,g(z)$, with the mass dependence described by a power-law and the redshift behavior described by a Gaussian function:
\be
\mathcal{S}(M,z)=\mathcal{S}_0\,M^\alpha\, \exp{\left[-\frac{(z-z_0)^2}{2\,\sigma_z^2}\right]}\;.
\label{eq:S}
\ee  
This completely defines our models. No additional ingredients, besides standard cosmological assumptions, will be introduced in the following. Then, in order to make predictions, we will only have to specify the values of the model parameters $\alpha$, $z_0$, and $\sigma_z$ (while $S_0$ will be set by the requirement of reproducing the RSB excess, as detailed below).

A different approach to predict the angular correlations of a specific population can be adopted if models of its $\de N/\de z$, luminosity function, and halo occupation are available. Our aim is instead to derive more general conclusions, without focusing on a given radio source population, and in this case, there would be three different functions and many more parameters to be arbitrarily defined. 
The idea behind the function used in Eq.~(\ref{eq:S}) is that typically the luminosity grows with the hosting halo mass up to a certain cutoff (in our case large masses will be effectively suppressed by the $\de n/\de M$) and that radio emissions are typically associated to star formation which starts at a certain $z_{\rm max}$ and possibly ends at a certain $z_{\rm min}$.
Despite the function used being an arbitrary choice, we will see that by varying its parameters, one can obtain very different number counts, luminosity functions, and bias, making it hard to imagine a physical scenario that is not represented, at least in an approximate way, in our sample.

For illustrative purposes, we will show plots for weak, medium and strong scalings with the mass by taking $\alpha$ equal to 0.1 (magenta), 0.5 (black) and 1 (cyan), and for populations with distribution in redshift peaked at low or intermediate $z$, or flat in $z$, by taking $(z_0,\sigma)$ equal to (0.3, 0.15) (dashed), (1, 0.5) (solid), and (0,+$\infty$) (dotted). We define the maximum redshift of the sources to be $z_\mathrm{max}=5$. The latter value is an arbitrary choice. It can be seen also as a physical assumption, since the most distant tracers of the radio source population, such as quasars, fall off beyond $z\sim 4$. However, results are insensitive to the choice.

The above models are chosen in order to cover a variety of physical scenarios. They are not intended to necessarily provide a good fit to anisotropy data (and most of them will not, as we will see later on), but to help understand the key physical requirements for a model to work. The values of the parameters that are preferred by our measurements will be derived in Section~\ref{sec:gal}.

The number counts are obtained through
\be
\frac{\de N}{\de \mathcal{S}}=\int \de z\,\frac{\de V}{\de z\,\de \Omega}\,\frac{\de n}{\de M\,\de V}\,\frac{\de M}{\de \mathcal{S}}\enspace,
\label{eq:dNdS}
\ee
where we use the notation for differential counts customary in the literature, i.e., without reporting $\de\Omega$ at the denominator. The cosmological volume factor is obtained from a $\Lambda$CDM model with parameters from \cite{Planck:2018vyg}. 
We consider halos between $M_{\rm min}=10^{6}\,M_\odot$ and $M_{\rm max}=10^{16}\,M_\odot$ (the precise values of the boundaries of the mass range do not affect our results in a significant way).
With the functional form of Eq.~\eqref{eq:S}, the last factor is just $\de M/\de \mathcal{S}=M/(\alpha\,\mathcal{S})$.  

The contribution to the radio background from extragalactic sources below the detection threshold is
\be
T_\mathrm{bts}=\frac{c^2}{2\,k_B\,\nu^2}\int^{\mathcal{S}_{\rm thr}}_0\,\de \mathcal{S}\frac{\de N}{\de \mathcal{S}}\,\mathcal{S}\;,
\label{eq:intens}
\ee
where we set $\mathcal{S}_{\rm thr}=5$ mJy for the LOFAR images considered in this work~\citep{Offringa:2021rwp,Cowie:2023rwp}.  For each model, i.e., each combination of $\alpha$, $z_0$ and $\sigma_z$, we obtain $\mathcal{S}_0$ in Eq.~\eqref{eq:S} by requiring $T_\mathrm{bts}=150$ K, namely, by normalizing each model to reproduce the RSB excess discussed in the Introduction.

We show the source counts of the nine models mentioned above in Fig.~\ref{fig:dNdS} (left panel). They span a variety of scenarios, with excess counts just below the threshold of detected sources (cyan) or peaking at much lower fluxes, with/without (black/magenta) a tail at intermediate fluxes (i.e., in ranges possibly accessible by near future telescopes).  

The expected auto-angular
power spectrum of the unresolved radio sources and the cross-spectrum with extragalactic tracers of the matter distribution can both be written as
\be
C_\ell^{\textrm{radio},i} = \int \de z\,\frac{1}{H(z)}\frac{W_\textrm{radio}(z)W_{\rm i}(z)}{\chi(z)^2}
P_{\textrm{radio},i}\!\!\left[k=\frac{\ell}{\chi(z)},z\right],
\label{eq:clgen}
\ee
where $\chi(z)$ is the comoving distance to redshift $z$, obeying $\de z/\de\chi=H(z)$ with $H(z)$ the Hubble rate, $W$ is the window function described below, and $P_{\textrm{radio},i}$ is the three-dimensional power spectrum. The subscript $i$ means that Eq.~\eqref{eq:clgen} is used both for computing the autocorrelation ($i=$radio) and cross-correlation with galaxy catalogs ($i=$galaxy) or CMB lensing ($i=$lensing).  For radio sources, the window function can be written as:
\be
W_\textrm{radio}(z)=\chi^2(z)\,\rho_s(z)\quad {\rm with}\quad \rho_s(z)=\int_{M_{\rm min}}^{M^\prime_{\rm max}}\de M\frac{\de n}{\de M\,\de V}\,\mathcal{S}(M,z)\enspace,
\label{eq:Wradio}
\ee
so again using the ingredients described at the beginning of this Section. The upper integration limit is $M^\prime_{\rm max}=\min[M_{\rm max},M(\mathcal{S}_{\rm thr})]$.
We show the window functions for our reference models in Fig.~\ref{fig:dNdS} (right panel), where one can also note that, since the mass dependence is integrated, see Eq.~\eqref{eq:Wradio}, there is very little dependence on $\alpha$.  The galaxy window function is described in Section~\ref{sec:gal}, and the lensing case in Section~\ref{sec:lens}.  The three-dimensional power spectrum $P_{\textrm{radio},i}$ can be computed in the framework of the halo model as the sum of two terms: $P=P^{1h}+P^{2h}$.  More details on the power spectrum for the different cases are provided in the next Sections.

\section{Radio Autocorrelation}
\label{sec:auto}

\begin{figure*}
\centering
\includegraphics[width=0.55\textwidth]{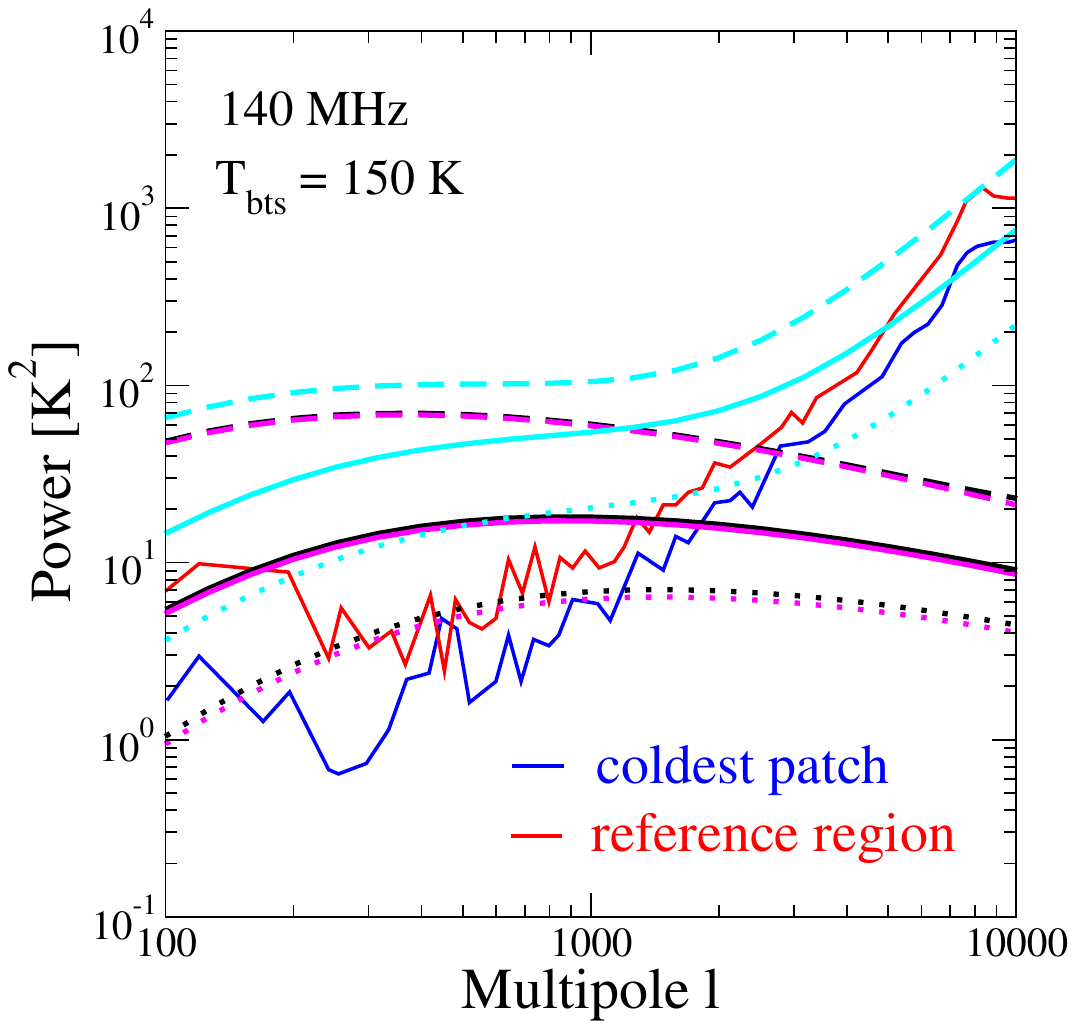}
\caption{The measured radio autocorrelation angular power spectrum obtained from the LOFAR images of the reference field (red) and coldest patch (blue) from~\protect\cite{Offringa:2021rwp}, along with autocorrelation angular power spectra that would result from the theoretical models of radio sources discussed in Sec. \ref{sec:mod}.  The color/style coding for the theoretical models is as in Fig.~\ref{fig:dNdS}, namely arising from Eq.~\eqref{eq:S} with weak, medium, and strong scalings with the mass by taking $\alpha$ equal to 0.1 (magenta), 0.5 (black) and 1 (cyan), and for populations with distribution in redshift peaked at low or intermediate $z$, or flat in $z$, by taking $(z_0,\sigma)$ equal to (0.3, 0.15) (dashed), (1, 0.5) (solid), and (0,+$\infty$) (dotted). }
\label{fig:APS}
\end{figure*}

In the case of the radio angular autocorrelation, we do not perform a statistical analysis, but nevertheless, we report an illustrative comparison of the predictions from the models considered in this work with the measurement in~\cite{Offringa:2021rwp}.

Assuming that the radio sources of our interest are point-like, the one-halo term of the power spectrum is given by
\be
P^{1h}(z)=\int^{M^\prime_{\rm max}}_{M_{\rm min}}\,\de M\ \frac{\de n}{\de M\,\de V} \,\left(\frac{\mathcal{S}(M)}{\rho_s(z)}\right)^2\;.
\ee
The assumption of point-like sources makes it $k$-independent.  The associated angular power spectrum is therefore flat in harmonic space and sometimes called Poisson-noise term $C_P$. Combining the above ingredients one gets
\be
C_P=\int \de z\,\frac{\de V}{\de z\,\de \Omega}\,\int_{M_{\rm min}}^{M^\prime_{\rm max}}\de M\frac{\de n}{\de M\,\de V}\,\mathcal{S}^2(M,z)=\int^{S_{\rm thr}}_0\,\de \mathcal{S}\frac{\de N}{\de \mathcal{S}}\,\mathcal{S}^2\;.
\label{eq:Cp}
\ee
Eq.~\eqref{eq:Cp} provides the dominant angular autocorrelation at very small scales.

The two-halo term is given by $P^{2h}(k,z)=\langle b_\textrm{radio} (z)\rangle^2 P_\textrm{lin}(k,z) $, where $P_\textrm{lin}$ is the linear matter power spectrum.
The radio effective bias is computed within the halo model approach, and averaging the halo bias $b_h[M,z]$
(taken from \cite{Sheth:1999mn}) with the radio flux distribution
\be
\langle b_\textrm{radio} (z)\rangle=\int^{M^\prime_{\rm max}}_{M_{\rm min}}\,\de M\ \frac{\de n}{\de M\,\de V} \,\frac{\mathcal{S}(M)}{\rho_s(z)}\times b_h[M(\mathcal{S}),z]\;.
\label{eq:biasradio}
\ee

In Fig.~\ref{fig:APS}, we show the autocorrelation of the source models compared with the measurement presented in \cite{Offringa:2021rwp}. 
Models with significant emission at low redshift are clearly disfavoured. The flat $z$ case (extending to $z_\mathrm{max}=5$) does not overshoot the data, even though some additional contribution might be needed to explain the rise of the measured power spectrum at $\ell\gtrsim 1000$. Despite no error bars are provided in Fig.~\ref{fig:APS}, we mention here that very high multipoles, $\ell\gtrsim 3000$ (which corresponds to $3.5^\prime$), refer to scales below the radio beam size and thus can suffer from significant uncertainty.
Note that since the 2-halo component of the 3D power spectra differ only for the bias term which is of order 1, and the radio window functions are nearly independent from $\alpha$, it is mostly the different redshift dependence that sets the amplitude of the predicted signal at large scales.

Let us also stress again that the quantitative analysis of this work will be focused only on the cross-correlation measurements presented below, which typically suffer from less systematic uncertainties than the autocorrelation.

\section{Cross Correlation between Radio Maps and CMB Lensing}
\label{sec:lens}
Now we move to a quantitative analysis based on cross-correlations, starting from the cross-correlation of the \textit{Planck}  CMB lensing convergence map with the LOFAR 140 MHz NCP mosaic image.  We use the NCP mosaic image instead of the individual LOFAR fields in order to maximize the area of overlap, so to enhance the cross-correlation signal. 

The path of CMB photons is deflected by the intervening gravitational potentials associated with the large-scale structure (see, e.g., \cite{lewis06} for a review).
This induces a remapping of the primary CMB anisotropies $X(\hat{\mathbf{n}})$ according to $\tilde{X}(\hat{\mathbf{n}}) = X\left(\hat{\mathbf{n}}+\nabla\phi(\hat{\mathbf{n}})\right)$, where $\tilde{X}(\hat{\mathbf{n}})$ are the observed CMB maps and $\phi(\hat{\mathbf{n}})$ is the CMB lensing potential.
These deflections, typically of a few arcminutes, introduce correlations between modes of the CMB anisotropies that can be exploited to reconstruct the projected gravitational potential $\phi(\hat{\mathbf{n}})$.
As a result, CMB lensing maps contain information on the integrated total matter distribution along the line-of-sight, with peak sensitivity around redshift $z \sim 2$ (see Fig.~\ref{fig:dNdS}). 
Conversely, radio galaxies are biased signposts of the same dark matter haloes that act as lenses for CMB photons.
It is therefore natural to expect a degree of correlation between maps of CMB lensing and radio data since they respond to the underlying dark matter field in complementary ways.
In fact, many statistically significant cross-correlation measurements between CMB lensing maps and galaxy catalogs extracted from radio surveys have been reported to date \citep[e.g.,][]{smith07,allison15,alonso21,piccirilli22}.
In this paper, we instead focus on the cross-correlation between CMB lensing convergence $\kappa = -\frac{1}{2}\nabla^2\phi$ and the spatial anisotropies of the radio background at 140 MHz.

The CMB lensing window function is given by~\cite{Bartelmann:2010fz}:
\be
W_\textrm{lensing}(\chi)=\frac{3}{2}\ho^2\om[1+z(\chi)]\chi\,\frac{\chi_*-\chi}{\chi_*}\;,
\label{eq:wcmb}
\ee
where $\ho$ is the Hubble constant today, $\om$ is the matter-density parameter, and $\chi_*$ is the comoving distance to the last-scattering surface, all taken from \cite{Planck:2018vyg}.
The shape of $W_\textrm{lensing}$ is shown in Fig.~\ref{fig:dNdS} (right). Note that it extends over a large range of redshifts.  The lensing bias is obtained by averaging the halo bias over the matter distribution
\be
\langle b_\textrm{lensing} (z)\rangle=\int^{M^\prime_{\rm max}}_{M_{\rm min}}\,\de M\ \frac{\de n}{\de M\,\de V} \,\tilde u(k|M,z)\times b_h[M,z]\;,
\label{eq:biaslens}
\ee
where $\tilde u(k|M,z)$ is the Fourier transform of the matter halo density profile.

The cross-correlation between radio data from LOFAR and \textit{Planck} CMB lensing is measured in harmonic space.  We use the pseudo-$C_\ell$ estimator implemented in the \texttt{NaMaster}\footnote{\url{https://github.com/LSSTDESC/NaMaster}} package \citep{alonso19} to extract the angular cross-power spectrum $C_\ell^{\kappa\Delta T}$. 
Given two masks to weight the observed maps, \texttt{NaMaster} calculates the mode-coupling matrix $M_{\ell\ell'}$ and undoes the effects of masking-induced statistical coupling between different harmonic modes.  To improve the condition number of the mode-coupling matrix, we bin all spectra into 15 linearly-spaced bandpowers between $50 \le \ell \le 2000$ (approximately corresponding to $\Delta\ell\simeq 139$) and return an unbiased estimate of the cross-power spectrum. We focus on $50 \le \ell \le 2000$ since the uncertainty becomes large outside this range (compared to model predictions) and there is no statistical gain in considering additional multipoles. When estimating the cross-power spectrum, we mask the lensing convergence map using the fiducial mask provided by the \textit{Planck} team which removes regions dominated by Galactic foregrounds as well as a number of extragalactic sources (including infrared galaxies and Sunyaev-Zel'dovich clusters). 
The LOFAR radio data are similarly weighted by a map of their sensitivity over the NCP mosaic image.
Due to the small size of the observed overlapping patch (approximately 100 deg$^2$), we work in the flat-sky limit and reproject the original \textit{Planck} data from the \texttt{HEALPix} \citep{healpix} pixelation scheme to the same LOFAR orthographic projection using the \texttt{reproject} package.\footnote{\url{https://reproject.readthedocs.io}}  We additionally correct the cross-power spectrum for the pixel window function, reprojection effects, and a small ($\sim 5\%$) multiplicative CMB lensing transfer function to account for the spatial-dependent normalization of the lensing map (see, e.g., \cite{carron22,alonso23}).  We calculate the covariance matrix of the band powers by cross-correlating 480 realizations of the CMB lensing convergence maps as reconstructed by the \textit{Planck} pipeline with the real LOFAR radio map.  This implicitly assumes the two maps to be uncorrelated, which is a well-founded assumption given the noise levels of the respective fields (especially CMB lensing).  We take the square root of the diagonal of the covariance matrix $\mathbb{C}_{\ell\ell'}$ as the statistical uncertainties on the recovered cross-power spectrum.

We do not detect a statistically significant cross-power spectrum.
Defining the null-hypothesis as the absence of correlation between the CMB lensing and the radio background, $C_\ell^{\kappa\Delta T}=0$, the chi-square value can be evaluated as $\chi^2_{\rm null} = \sum_{\ell\ell'} \hat{C}^{\kappa\Delta T}_\ell \mathbb{C}^{-1}_{\ell\ell'}\hat{C}^{\kappa\Delta T}_\ell \simeq 19.6$, corresponding to a probability-to-exceed of $12\%$.\footnote{When inverting the covariance matrix $\mathbb{C}_{\ell\ell'}$, we account for biases due to the limited number of simulations by applying the Hartlap factor, a correction that adjusts the inverse matrix for enhanced accuracy \citep{Hartlap:2006kj}.} Therefore, we cannot rule out the null hypothesis.
A comparison between our reference models and the measurement is provided in Fig.~\ref{fig:CCF_SDSS} (left).  Note that, since the CMB lensing window function is rather flat in redshift, the predictions from the different models are rather similar (remember they have the same overall ``normalization" given by matching Eq.~\eqref{eq:intens} with $T_\textrm{RSB}$). This highlights the importance of this cross-correlation which has the potential to constrain different interpretations of the RSB excess with little dependence on the specific model. On the other hand, the data in Fig.~\ref{fig:CCF_SDSS} are poorly constraining.  A measurement with a larger radio field, so to increase the statistics and reduce errors, is in order to fully exploit this signature for constraining the RSB excess.

\section{Cross Correlation between Radio Maps and Galaxy Catalogs }
\label{sec:gal}

\begin{figure*}
\vspace{-3cm}
\centering
\includegraphics[width=0.51\textwidth]{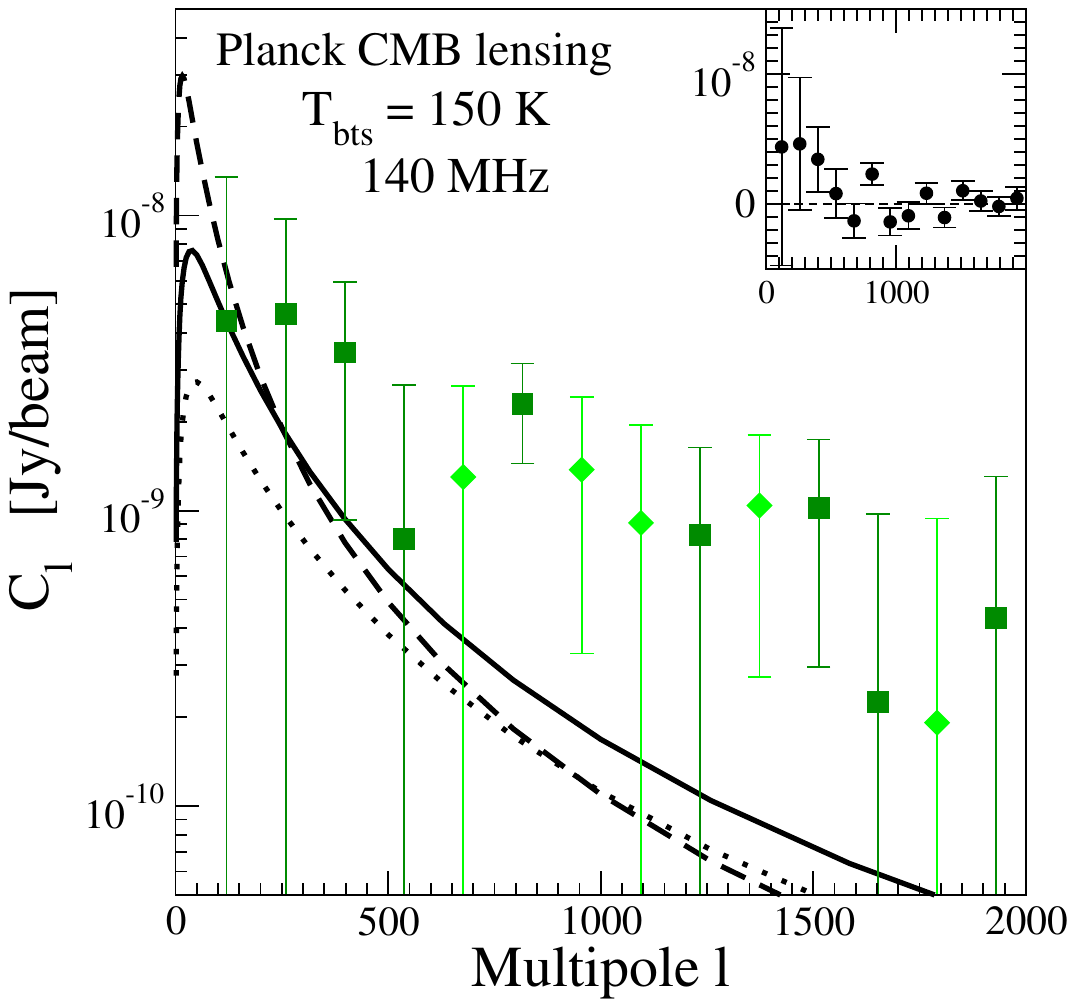}
\includegraphics[width=0.48\textwidth]{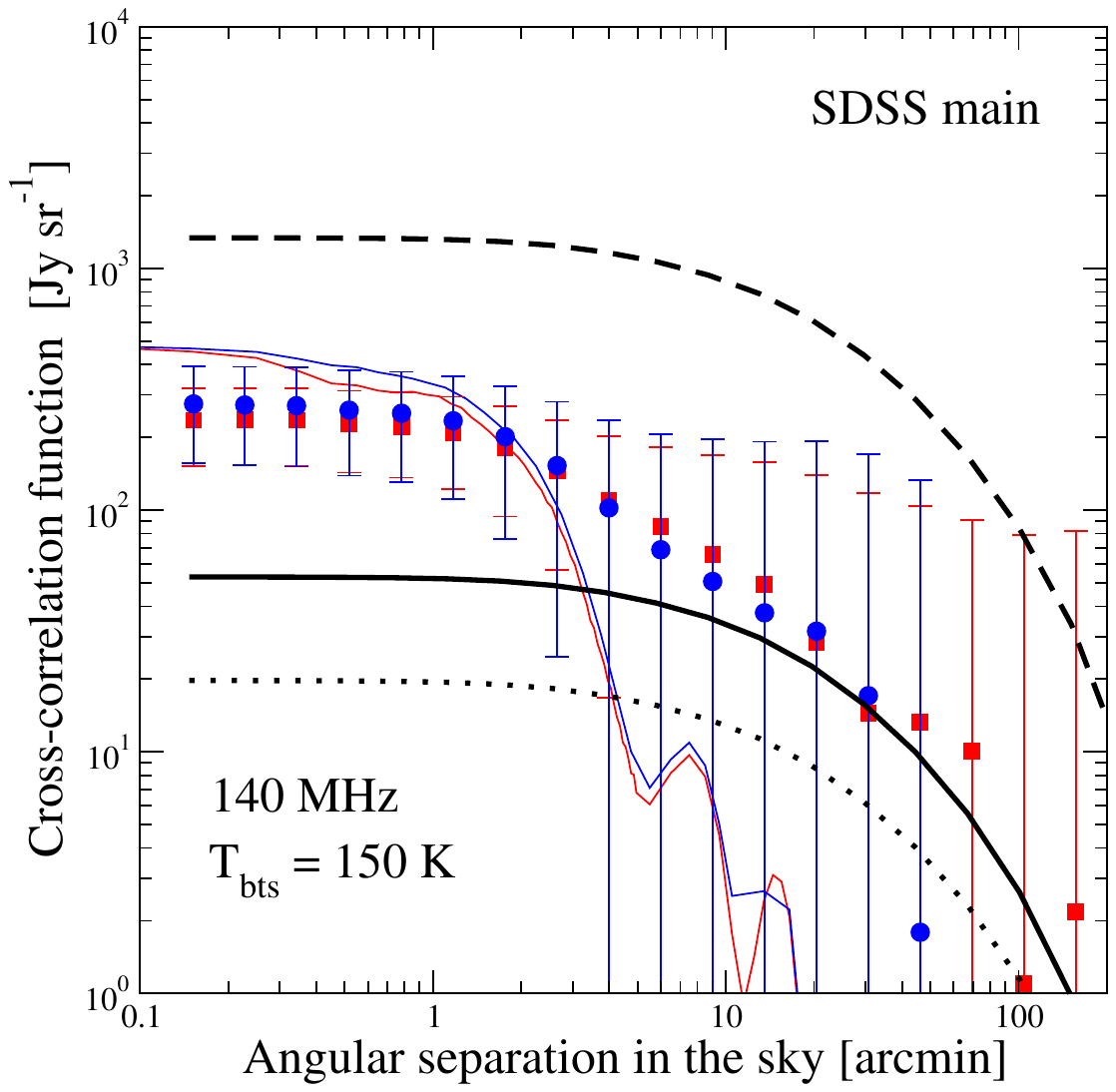}
\caption{Left: Measured cross-correlation angular power spectrum between the LOFAR NCP mosaic image and CMB lensing measured by \textit{Planck}. Black lines are the predictions for the models described in the text (including only the clustering, i.e, 2-halo, term), with style coding as in Fig.~\ref{fig:dNdS}, namely for populations with distribution in redshift peaked at low or intermediate $z$, or flat in $z$, by taking $(z_0,\sigma)$ equal to (0.3, 0.15) (dashed), (1, 0.5) (solid), and (0,+$\infty$) (dotted). Green points are the measurements at different multipoles with dark green for positive values and light green for negative ones. For illustrative purposes, we show the measurement also in linear scale in the inset plot. 
Right: Measured cross-correlation angular power of the LOFAR images of the reference field (red) and coldest patch (blue) with SDSS main catalog.  Solid blue and red lines show the LOFAR radio beams in the two fields. Black lines are the predictions for the same models as in the left panel.}
\label{fig:CCF_SDSS}
\end{figure*}

In this Section, we analyze the cross-correlation of the RSB with galaxy catalogs.
Contrary to the CMB lensing case, where we used the NCP mosaic image, here we consider the coldest patch and reference individual LOFAR fields. Despite having a smaller area, they are more suitable for the purposes of this Section, i.e., they provide a more significant signal, because of the lower noise of the images and since the NCP location of the mosaic image is not ideal for galaxy surveys.

In the case of galaxies, the window function entering Eq.~\eqref{eq:clgen} simply reads $W_\textrm{galaxies}(z)=H(z)\,dN_g/dz$, where $dN_g/dz$ is the redshift distribution of galaxies (normalized to unity).
The window functions used in this work are derived directly from the redshift of the sources reported in the catalogs. They are shown with black solid lines in Fig.~\ref{fig:dNdS} (right panel).

At the scale of interest, and considering sufficiently small systems (i.e., galaxy sizes), the one-halo term of the 3D power spectrum is $k$-independent, leading to an $\ell$-independent angular power spectrum. This might be not the case for cluster emission and very low redshifts (as the ones in the 2MPZ catalog). However, an explanation of the RSB excess in terms of clusters is viable only if we assume they are surrounded by a giant halo which would be at least partially resolved out by the LOFAR interferometer (and beyond extended emissions detected so far around clusters). On the contrary, if we assume their size is such that they are fully detected in LOFAR observations, clusters are too rare to offer a significant contribution to the RSB excess. Therefore, with the data considered in this work, we do not evaluate clusters and we can simplify our treatment assuming point-like sources, i.e., $C_\ell^{1h}=$const. Since the estimate of $C_\ell^{1h}$ heavily depends on details of the population under investigation, i.e., on the matching between the type of galaxies in the catalog under consideration and the source of radio emission, the prediction of the cross-correlation signal is highly model-dependent, i.e., it would require a precise estimate of the halo occupation distribution of both galaxies and radio emitters. Given the spirit of this work of deriving general conclusions, we decided to leave the effective shot noise level, i.e., $C_\ell^{1h}$, as a free parameter, determined by the fit.
To better understand this point, consider, e.g., two radio source populations providing the same RSB, but in one case given by a ``few" bright sources and in the other by many dim sources. In the first case, if all the sources are hosted by the galaxies of the catalog involved in the cross-correlation, then the one-halo term would be large. On the contrary, if they are hosted by different galaxies (so do not belong to the same halo), there is no one-halo correlation. In the case of the numerous and faint population, 
they are distributed in so many halos that a single galaxy catalog includes only a fraction of them and in turn, the associated one-halo term would be small.  Given the awkward link between the RSB contribution and the correlation, the one-halo term is not the main subject of this work.

The two-halo term can be instead computed in a much more reliable way and it is given by $P^{2h}_{\textrm{radio},\textrm{gal}}(k,z)=\langle b_\textrm{gal} (z)\rangle\,\langle b_\textrm{radio} (z)\rangle P_\textrm{lin}(k,z) $, with the effective bias of galaxies taken from the literature: for 2MPZ we follow \cite{Balaguera-Antolinez:2017dpm}, for SDSS main \cite{Repp:2020kfd}, and for SDSS quasars \cite{Laurent:2017gze}.

We express the results of the cross-correlation in terms of the correlation function $\xi$, instead of the angular power spectrum $C_\ell$, the two being related by the transformation:
\be
\xi(\theta)=\sum_\ell \frac{2\ell+1}{4\pi} \tilde C_\ell\,P_\ell(\cos\theta)\;,
\label{eq:CCF}
\ee
where $P_\ell$ are the Legendre polynomials, and $\tilde C_\ell=C_\ell\,\tilde W_\ell^\textrm{radio}$ with $C_\ell$ from Eq.~\eqref{eq:clgen} and $\tilde W_\ell^\textrm{radio}$ being the radio beam window function (we neglect the galaxy beam since it is much smaller than the radio one). 

We already saw that the dependence on $\alpha$ was mild in the case of autocorrelation. It is even milder for the cross-correlation, where the bias enters just linearly (instead of quadratically as in the autocorrelation). The illustrative plots of this Section will refer only to the case with $\alpha=0.5$.

\begin{figure*}
\centering
\includegraphics[width=0.48\textwidth]{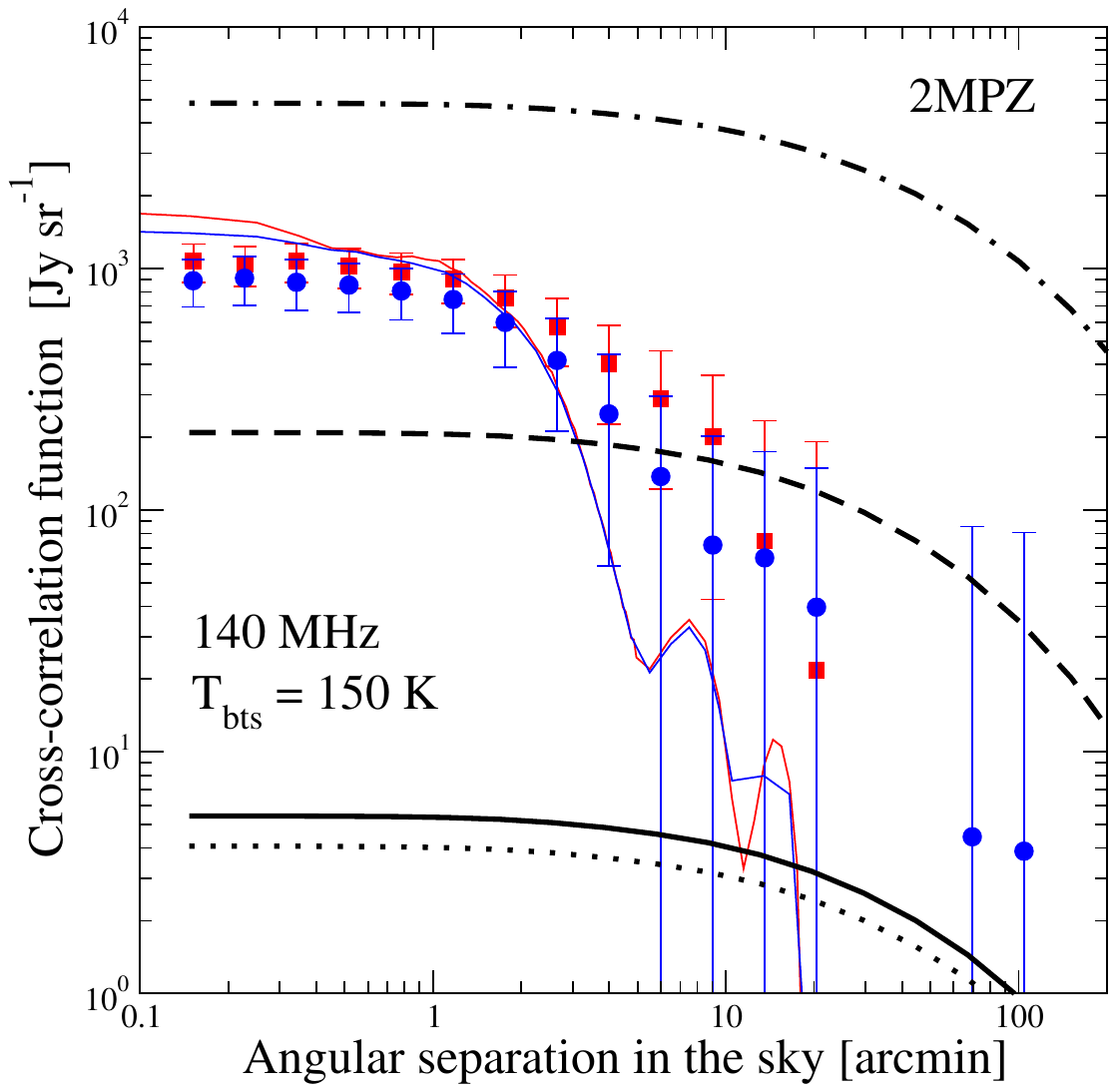}
\includegraphics[width=0.48\textwidth]{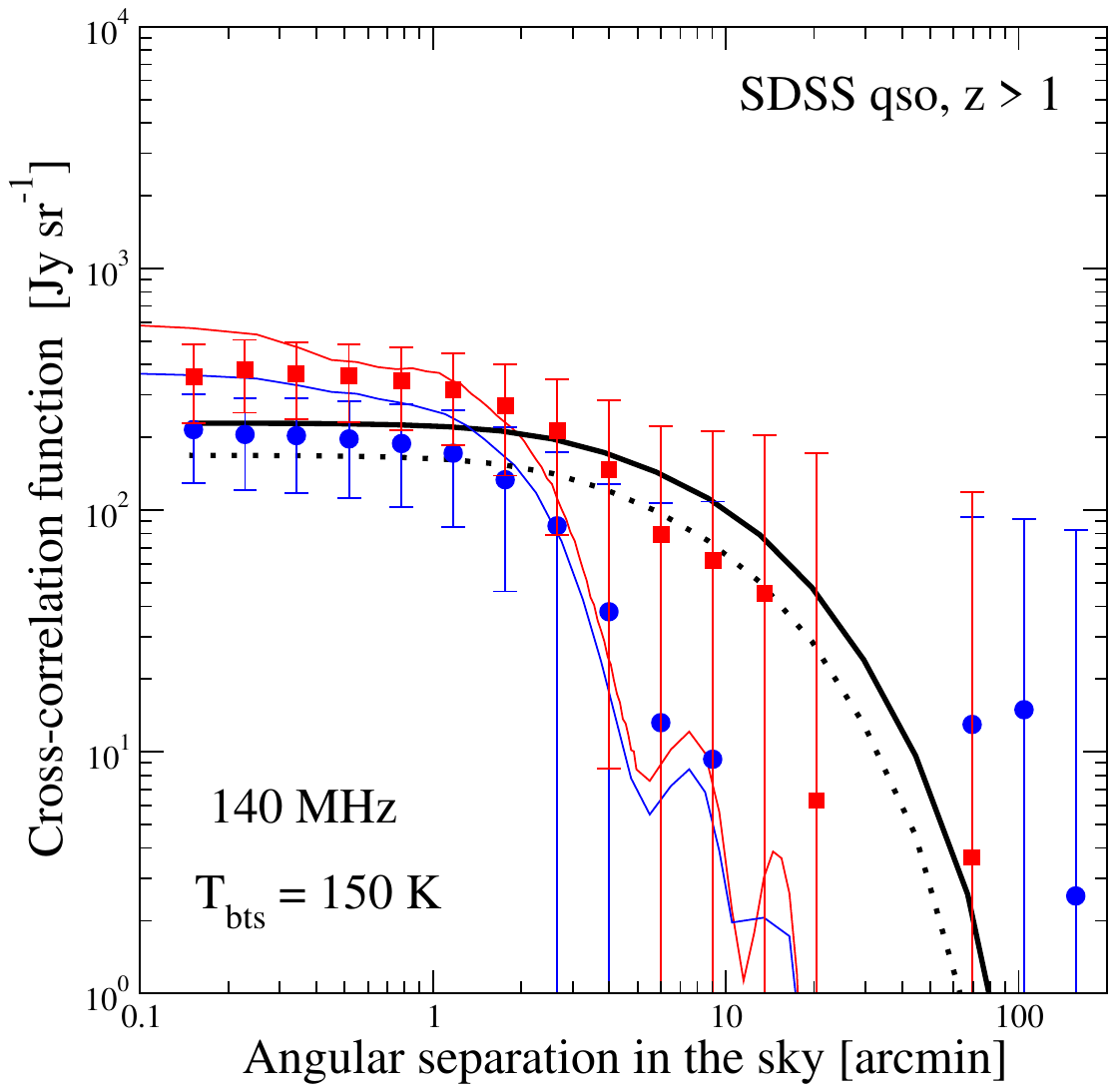}
\caption{Measured cross-correlation angular power of the LOFAR images of the reference field (red) and coldest patch (blue) with the 2MPZ (left) and SDSS quasar (right) catalogs, with error bars. Color/style coding as in Fig.~\ref{fig:CCF_SDSS} with $(z_0,\sigma)$ equal to (0.3, 0.15) (dashed), (1, 0.5) (solid), and (0,+$\infty$) (dotted), with the addition of a model with $(z_0,\sigma)= (0.1, 0.05)$ (dashed-dotted line) in the left panel to emphasize that this measurement excludes very low-$z$ explanations of the RSB excess.  Solid blue and red lines show the LOFAR radio beams in the two fields.
}
\label{fig:CCF_2}
\end{figure*}

We perform the measurement of the cross-correlation between LOFAR images at 140 MHz and galaxy catalogs described in Section~\ref{sec:obs} using the ``NKCorrelation" class from the TreeCorr package~\citep{treecorr}. This class handles two-point correlation functions between a catalog of source positions and a scalar quantity, in our case the radio flux in each pixel of the image under consideration.
The cross-correlation is computed in 20 logarithmically spaced bins, with centers ranging from 0.12 to 420~arcmin. The correlation estimator $\xi(\theta)$ is defined such that
$\int d\theta\, \xi(\theta) = 0$.\footnote{In the measurement, this condition refers to an integral up to the largest measured scale, whilst, in the theoretical model, the integral is over the whole sky. Since our correlation analysis is restricted to separations smaller than the size of the observed patch, the difference is negligible and can be safely ignored.}
The covariance matrix is then estimated using the jackknife method with 300 patches.

The measured cross-correlation of the LOFAR images with SDSS main catalog is shown in Fig.~\ref{fig:CCF_SDSS} (right) and with the 2MPZ and SDSS quasar catalogs in Fig.~\ref{fig:CCF_2}. We show the predicted cross-correlation for the different models with black lines. We also superimpose a Poisson-noise term, i.e., a contribution with $C_\ell=$const, (red and blue solid lines), arbitrarily normalized, since, as already mentioned, our modeling aims to be very general but cannot precisely predict the one-halo component (i.e., the correlation at angular size $\lesssim$ radio beam size). Note that a theoretical term $C_\ell=$~const has to be convolved with the telescope beam to be compared with observations, and therefore when transformed to real space it has exactly the beam shape.

It is clear that low redshift scenarios, such as the case with $(z_0,\sigma)= (0.3, 0.15)$, are excluded by SDSS main data. In Fig.~\ref{fig:CCF_2} (left), we add a case with  $(z_0,\sigma)= (0.1, 0.05)$, shown with dashed-dotted line, to illustrate that 2MPZ is very effective in excluding very low-$z$ models.  Therefore, as for the case of autocorrelation, we find that the low-$z$ cases are challenged by data.  High redshift scenarios are instead compatible with our cross-correlation measurements. 

\begin{figure*}[th]
\centering
\includegraphics[width=0.45\textwidth]{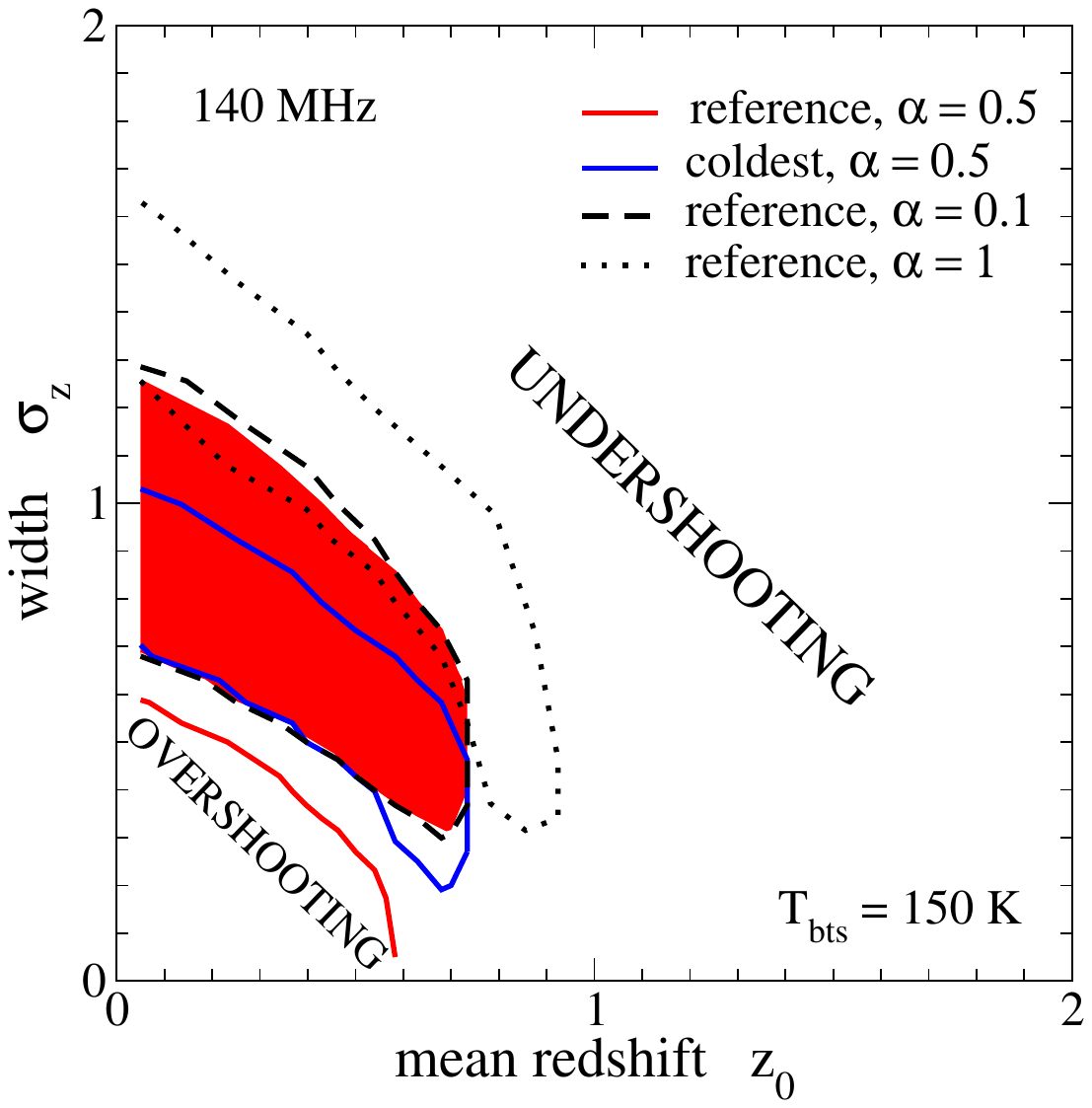}
\includegraphics[width=0.45\textwidth]{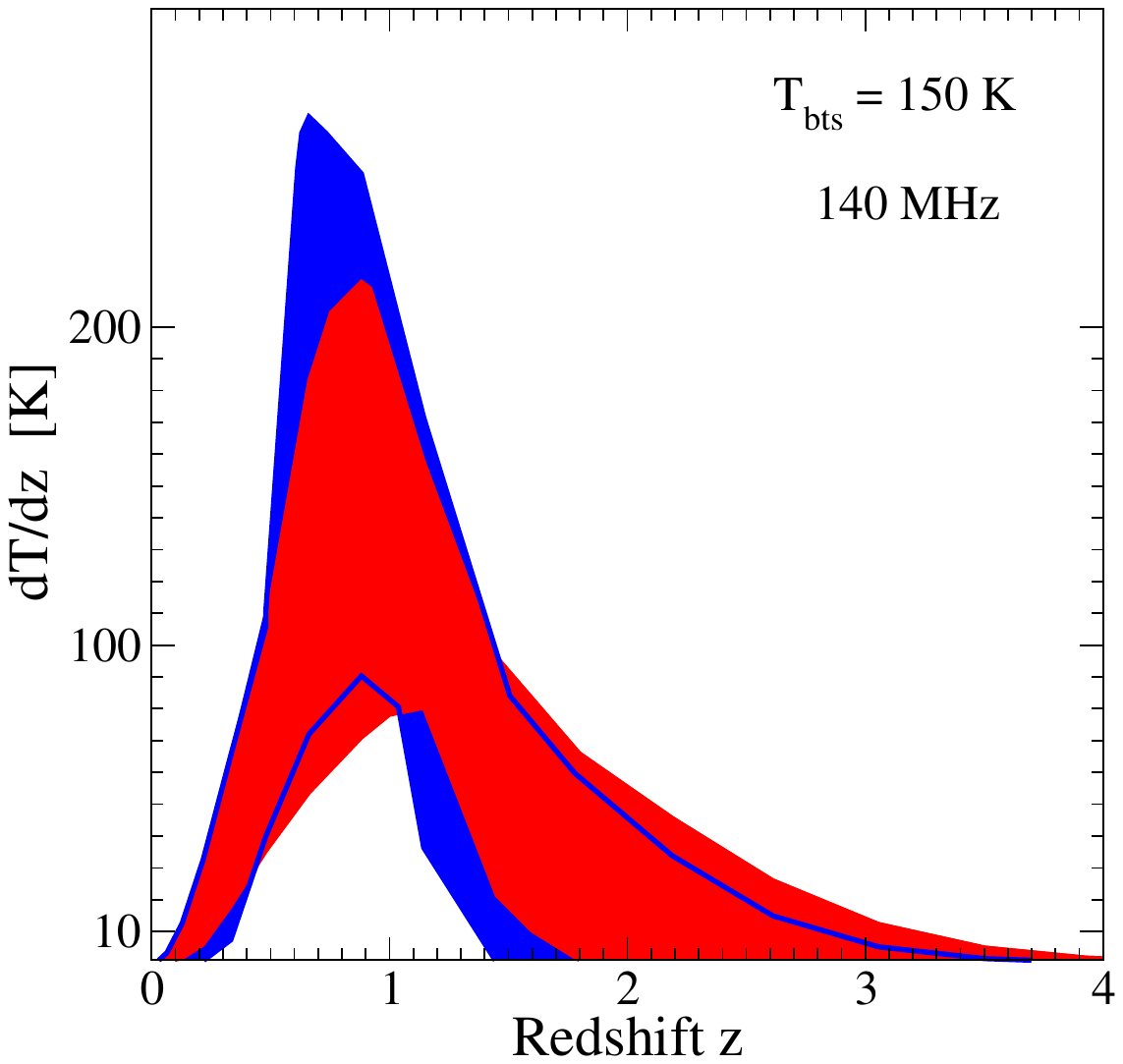}
\caption{Left: Closed contours are the 99\% C.L. regions of the two parameters $(z_0,\sigma)$ describing the redshift behaviour of radio sources. The regions are obtained by fitting the cross-correlation measurements shown in Figs.~\ref{fig:CCF_SDSS} and \ref{fig:CCF_2} for the reference LOFAR field (red filled region for $\alpha=0.5$ and black line for $\alpha=0.1$ (dashed) and $\alpha=1$ (dotted)) and coldest patch (blue line). The red solid line shows the $8\sigma$ (open) contour for $\alpha=0.5$ and the reference LOFAR field. Right: Contribution to the extragalactic radio background as a function of the redshift, taking the envelope of all models belonging to the preferred regions (see left panel), with $\alpha\in [0.1,1]$ and for reference LOFAR field (red) and coldest patch (blue).}
\label{fig:summary}
\end{figure*}

To be more quantitative about this last statement and about whether a scenario involving spatial correlation of the RSB origin population with that of galaxy catalogs is preferred to a scenario without such spatial correlation, we perform a global fit including all four cross-correlation measurements.  Six free parameters ($z_0,\,\sigma_z,\,C^{1h}_{2MPZ},\,C^{1h}_{SDSSm},\,C^{1h}_{SDSSq}\,C^{1h}_{lens}$) are present in the fit. We set $\alpha$ to three different cases $\alpha=0.1,\,0.5,\,1$. The $\chi^2$ is computed using the covariance matrix discussed above.\footnote{We also note that the covariance matrix is highly non-diagonal, so conclusions drawn by looking directly at the plots, where error bars are reported from the diagonal of the covariance matrix, have to be verified.}
In Fig.~\ref{fig:summary} (left), we show as a red-filled area the 99\% C.L. region of ($z_0,\,\sigma_z$) obtained after profiling out the four one-halo parameters $C^{1h}_{i}$.
The region at low ($z_0,\,\sigma_z$), i.e., below the preferred region, overshoots low-$z$ measurements and is excluded.  The region at large ($z_0,\,\sigma_z$), i.e., above/right of the preferred region, undershoots the measurements. So, a source population peaked at high-z would not explain the SDSS main measurement, but, at the same time, it is also not constrained by our analysis and could contribute to the RSB.

We note here that cases with $\alpha\lesssim 0.5$ provide nearly identical contour regions. For larger values, $\alpha\simeq 1$, the contribution from massive halos increases and since they are more numerous at low redshift than at high redshift, the bounds become somewhat more constraining (see dotted line with respect to dashed line and red band).  The two LOFAR fields, reference (red) and coldest (blue) provide compatible results, with the latter being slightly more constraining, as expected.

To better understand whether or not we need an extragalactic component from large-scale clustering to fit the individual cross-correlation measurements of galaxies with the LOFAR fields, we compare a model with only a (beam-like) Poisson-noise term with a model where we also add the two-halo component coming from clustering. For the latter term, since the angular dependence of the correlation function is not dramatically different for different models, see Figs.~\ref{fig:CCF_SDSS} and \ref{fig:CCF_2}, we consider the case with $\alpha=0.5$ and $(z_0,\sigma)= (1.0, 0.5)$, with a parameter providing the overall normalization and determined by the fit.
Thus we have one normalization parameter for the beam-like model (which can be seen as parametrizing our ignorance about the shot noise contribution) and two parameters when adding the 2-halo term (the second one essentially parametrizing our ignorance about the clustering bias).
The $\Delta \chi^2$ can be thus taken with one degree of freedom. Values are reported in Table~\ref{tab:2h}.

\begin{table}
\centering
 \begin{tabular}{||c | c | c | c ||} 
 \hline
 $\Delta \chi^2$ & 2MPZ & SDSS main & SDSS qso   \\ [0.5ex] 
 \hline\hline
 reference & 6.2 & 49.7 & 8.0 \\ 
 \hline
 coldest & 0.5 & 46.0 & 0.3 \\
 \hline
\end{tabular}
\caption{$\Delta \chi^2$ values obtained from $\chi^2_{PN}-\chi^2_{PN2h}$, where in the model providing $\chi^2_{PN}$ only a Poisson-noise term scaling as the radio beam is included in the fit, whilst for $\chi^2_{PN2h}$ we add a clustering term, computed from the 2-halo correlation described in the text.}
\label{tab:2h}
\end{table}

The cross-correlation with the SDSS main catalog shows clear evidence for an extragalactic clustered component. The fact that we obtain closed contours in Fig.~\ref{fig:summary} (left), is mostly due to the correlation with the SDSS main catalog. The presence of a two-halo term in the cases of 2MPZ and SDSS quasar is more uncertain.
Table~\ref{tab:2h} shows that the evidence of the signal is at a statistical significance
$\gtrsim 7\sigma$.  This implies that at least a fraction of the RSB must be of extragalactic origin in order to provide such a correlation.  We estimate the minimum RSB fraction required, by considering a radio population with a window function matching the one of the SDSS main catalog, shown in Fig.~\ref{fig:dNdS}, imposing its cross-correlation to be compatible at 99\% C.L. with our measurement, and exploring different bias obtained from $\alpha\le$1. Then we compute $T_{\rm bts}$ from Eq.~\eqref{eq:intens} and find values always larger than 20\% of the RSB excess. In other words, we conclude that, in order to explain the cross-correlation of the LOFAR images with the SDSS main catalog, at least 20\% of the RSB at 140 MHz has to come from an extragalactic and clustered source population. 

In contrast, we see that the RSB excess cannot predominantly come from sources at low redshift.  On the left panel of Fig.~\ref{fig:summary} we draw the $8\sigma$ contour (solid red line) which is indeed an open contour, excluding only models at low redshift but not the high redshift ones.

In Fig.~\ref{fig:summary} (right), we summarize the fraction of the RSB excess that can come from a certain redshift, by plotting $\de T/\de z$ as a function of $z$ for all the models within the 99\% C.L. region of the profiled $(z_0,\sigma)$ distribution (i.e., within the regions like the ones reported in the left panel for $\alpha=0.1,\,0.5,\,1$ in the case of the reference field and for $\alpha=0.5$ for the coldest patch). The result is shown considering $\alpha\in [0.1,1]$.
The expression of $\de T/\de z$ can be computed combining Eqs.~\ref{eq:dNdS} and \ref{eq:intens}:

\be
\frac{\de T}{\de z}=\frac{c^2}{2\,k_B\,\nu^2}\int^{\mathcal{S}_{\rm thr}}_0\,\de \mathcal{S}\,\frac{\de V}{\de z\,\de \Omega}\,\frac{\de n}{\de M\,\de V}\,\frac{\de M}{\de \mathcal{S}}\, \mathcal{S}\;.
\label{eq:dTdz}
\ee
By taking at the upper boundary of the red band, we find that the contribution for $z\lesssim 0.5$ amounts to $\lesssim 15\%$ of the RSB. 

One might wonder if the detected correlation can be explained by known source populations, i.e., by their components below the detection threshold. Using the model of counts in \citet{Intema:2017}, we found that the RSB contribution from below threshold sources amounts to 2 K. The counts essentially refer to AGNs. At 140 MHz the number counts of star-forming galaxies (SFGs) has been not determined. We can however use information from deeper surveys at higher frequencies to estimate it. From the model in \citet{Gervasi:2008rr}, we found an RSB contribution from SFGs of about 10 K at 140 MHz. Therefore the overall contribution from the unresolved components of known sources is $\sim 10\%$ of the total RSB, not far from the 20\% mentioned above. The redshift behavior plotted in Fig.~\ref{fig:summary} is compatible with an SFG population. Therefore, considering that the extrapolation of the counts is an oversimplified estimate, it is not inconceivable that sub-threshold sources of known type could account for the measured cross-correlation.

\section{Conclusion}
\label{sec:con}

We have investigated the angular cross-correlation power spectra of LOFAR images of the diffuse radio emission with galaxy and quasar catalogs from SDSS and with the map of CMB lensing from Planck.
We have considered the cross-correlation with source-subtracted radio maps, as opposed to radio sources extracted from maps, with the goal of constraining extragalactic origin scenarios for the RSB.  

As discussed in \S \ref{sec:lens}, we do not detect any significant correlation between the radio images and CMB lensing, likely in part because of the limited region of effective overlap between the two maps.  Still, since the CMB lensing extends to a wide redshift range and can overlap with very different RSB interpretations, we believe that cross-correlations of CMB lensing with radio maps are a useful avenue to pursue further.

As discussed in \S \ref{sec:gal}, we find that models of the RSB resulting from very low redshift sources ($z \sim$ 0.1-0.3) greatly overproduce the measured cross-correlation power spectrum between the LOFAR radio maps and galaxy catalogs, excluding those redshifts of origin for the predominant source classes of the RSB.  However, models of the RSB resulting from intermediate or higher redshift sources ($z \gtrsim$0.5) are not incompatible with the measured cross-correlation power.  This is true even for a wide range of faint radio source count ($dN/dS$) distributions.  Generally, as discussed in \S \ref{sec:gal}, a scenario involving spatial correlation of the RSB origin population with that of the SDSS main galaxy catalog is preferred to a scenario without such spatial correlation, favoring hypothetical source populations that trace the matter distribution in the universe.  In \S \ref{sec:gal}, we calculate that at a minimum 20\% of the RSB surface brightness level must originate from populations tracing the large-scale distribution of matter in the universe, indicating that at least this fraction of the RSB is of extragalactic origin.

These results are compatible with the analyses in \cite{Offringa:2021rwp} and \cite{Cowie:2023rwp}, which both showed an excess of radio anisotropy power above that which would result from known extragalactic radio source classes.  The latter work did raise the possibility of significantly extended sources as an important source class which this analysis does not constrain. Here we see that if 100\% of the RSB is produced by a new class of numerous, low-flux radio sources, these sources should be at intermediate or higher redshifts. This class of sources has to provide a strong evolution of the far infrared-radio correlation~\citep{Ysard:2012pw} so as not to overproduce the measured level of the far infrared background. Whether this can be compatible with constraints derived for known source classes (see, e.g., \citep{Magliocchetti:2022xmz}) depends on the redshift distribution of their radio emission. Looking forward, in order to ascribe the RSB surface brightness excess to a specific particular extragalactic population with angular cross-correlation analyses, it would be crucial to measure a cross-correlation at $z\sim 1$ and angular scales much greater than radio beam size.

\section*{Acknowledgements}
MR and ET acknowledge support from the project ``Theoretical Astroparticle Physics (TAsP)'' funded by the INFN, from `Departments of Excellence 2018-2022' grant awarded by the Italian Ministry of Education, University and Research (\textsc{miur}) L.\ 232/2016, from the research grant `From Darklight to DM: understanding the galaxy/matter connection to measure the Universe' No.\ 20179P3PKJ funded by \textsc{miur}, and from the ``Grant for Internationalization" of the University of Torino. F.B. acknowledges support by the Department of Energy, Contract DE-AC02-76SF00515. The work of S.H.~is supported by the U.S.~Department of Energy Office of Science under award number DE-SC0020262, NSF Grant No.~AST1908960 and No.~PHY-2209420, and JSPS KAKENHI Grant Number JP22K03630 and JP23H04899. This work was supported by World Premier International Research Center Initiative (WPI Initiative), MEXT, Japan.

\section*{Data Availability}
The data underlying this article will be shared on reasonable request
to the corresponding authors.



\bibliographystyle{mnras}

\bibliography{refs}

\appendix
\section{Redshift distribution estimation with Tomographer}\label{sec:tomo}

In this Appendix, we further validate our main analysis by inferring the unknown redshift distribution of the source population contributing to the radio background using the publicly available \texttt{Tomographer} tool.\footnote{Available at \url{http://tomographer.org}}
\texttt{Tomographer} is a web-based platform developed by Yi-Kuan Chiang and Brice Ménard that uses the clustering-based redshift algorithm with the approach in proposed in \citet{Menard:2013aaa} and the specific implementation of \citet{Chiang:2018miw}.
In a nutshell, the clustering redshift method enables the inference of redshift distributions from measurements of the spatial clustering of arbitrary sources (or diffuse intensity maps), using a set of reference objects for which redshifts are known (see, e.g., \citet{Newman:2008mb,Menard:2013aaa,Rahman15,Chiang:2018miw} for a thorough discussion of the method's principles and validation).
The key idea hinges on the spatial clustering observed on the sky when matter tracers share overlapping redshift distributions (neglecting gravitational lensing effects). 
In the case of \texttt{Tomographer}, the spectroscopic reference sample is constructed from a compilation of about two million Sloan Digitial Sky Survey sources (including galaxies and quasars) distributed over approximately 10,000 deg$^2$ and extending out to redshift $z \lesssim 4$.

Given a sky map (or a source catalog), \texttt{Tomographer} returns a measurement of the bias-weighted redshift distribution of the photons contributing to the radio background as well as an estimate of its uncertainties (estimated through a resampling approach).
More specifically, \texttt{Tomographer} computes the spatial correlation function $w(\theta)$ between the pixelized density maps (at an \texttt{HEALPix} resolution of $N_{\rm side} = 2048$) of the reference spectroscopic sample and of the input diffuse map (or source catalog). To enable a fast calculation of the results,  the redshift and angular binnings are pre-set to allow the algorithm to use a number of pre-computed quantities. 
Schematically, the cross-correlation amplitude $\bar{w}_{IR}$ between the input data $I$ and the reference data $R$ calculated in the reference redshift bin $z_i$ is given by 
\begin{equation}
    \label{eq:tomographer}
    \bar{w}_{IR}(z_i) = \frac{dI}{dz}(z_i) b_I(z_i)b_R(z_i)\bar{w}_{\rm DM}(z_i),
\end{equation}
where $b_I$ and $b_R$ are the effective linear bias parameters for the input/reference data, $\bar{w}_{\rm DM}$ is the dark matter clustering, and $\frac{dI}{dz}(z_i)$ is the quantity of interest. 
Note that the correlation function amplitudes $\bar{w}$ are obtained by integrating the correlation functions $w(\theta)$ over the angles that correspond to physical scales $2 \le r \le 8$ Mpc and that the scales larger than 2 deg are filtered out to mitigate potential large-scale systematics (including stellar contamination and dust extinction). 
By looking at Eq.~\ref{eq:tomographer}, we can see how the ratio between the cross-correlation and reference sample auto-correlation is sensitive to the bias-weighted redshift distribution of the unknown sample, $\bar{w}_{IR}/\bar{w}_{RR}\propto \frac{dI}{dz}\times b_I$.

\begin{figure}
\centering
\includegraphics[width=0.45\textwidth]{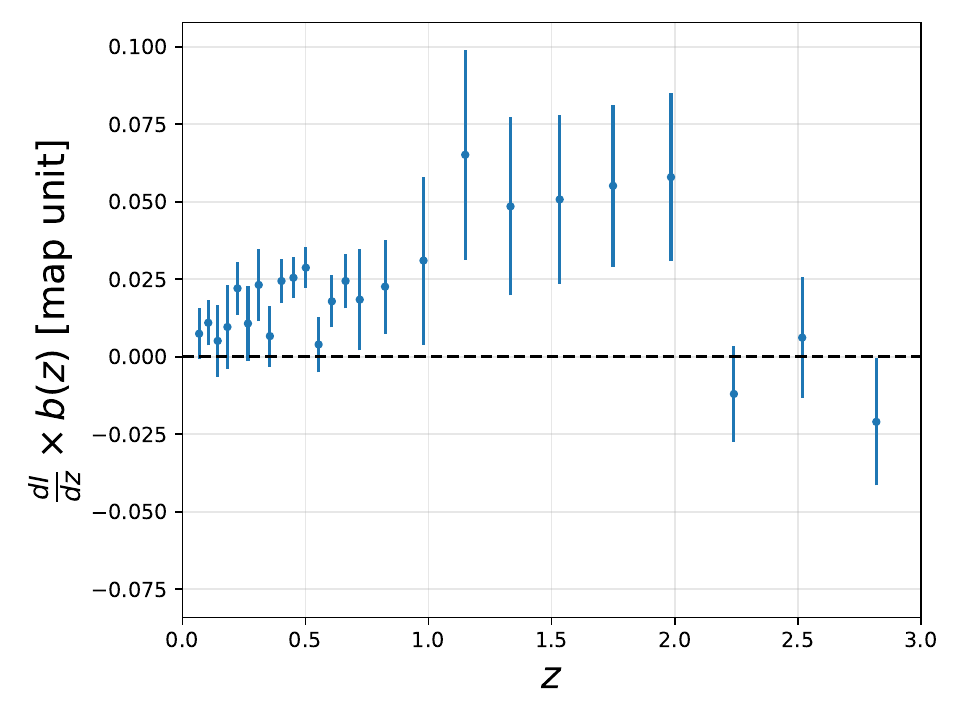}
\caption{Inferred bias-weighted redshift distribution of the photons contributing to the radio background in the LOFAR coldest patch. The units are the same ones of the original LOFAR map, i.e. Jy/beam.}
\label{fig:tomographer}
\end{figure}

In Fig.~\ref{fig:tomographer}, we show the recovered $\frac{dI}{dz}\times b(z)$ estimated from the LOFAR coldest patch. 
As can be seen, the inferred distribution qualitatively agrees with that shown in Fig.~\ref{fig:summary}, exhibiting a bulk contribution from structures at $z$ around 1, once a (moderate) bias evolution with redshift is foreseen.

\label{lastpage}
\end{document}